\documentclass[fleqn,usenatbib]{mnras}

\usepackage{newtxtext,newtxmath}

\usepackage[T1]{fontenc}

\DeclareRobustCommand{\VAN}[3]{#2}
\let\VANthebibliography\thebibliography
\def\thebibliography{\DeclareRobustCommand{\VAN}[3]{##3}\VANthebibliography}

\usepackage{graphicx}	
\usepackage{amsmath}	
\usepackage{dsfont}
\usepackage{subcaption} 
\usepackage{orcidlink}
\usepackage[normalem]{ulem}
\usepackage{xcolor}
\newcommand\removetexttmp{\bgroup\markoverwith{\textcolor{red}{\rule[0.5ex]{2pt}{1.4pt}}}\ULon}

\def\deltalin{\delta^{\mathrm{lin}}}
\def\deltanl{\delta^{\mathrm{nl}}}
\def\d{\mathrm{d}}
\newcommand{\condprob}[2]{\mathbb{P}\left({#1} \Bigm\vert {#2} \right)}
\newcommand\Eqn[1]{Eq.\,(\ref{#1})}
\newcommand\Eqns[2]    {Eqs.\,(\ref{#1}) and~(\ref{#2})}
\newcommand\Fig[1]     {Fig.\,{\ref{#1}}}
\newcommand\Tab[1]     {Table~\ref{#1}}

\newcommand\App[1]     {Appendix~\ref{#1}}
\newcommand\Sect[1]    {\S\ref{#1}}
\newcommand{\bracmat}[4]{\left[\begin{array}{c|c} {#1} & {#2} \\ \hline {#3} & {#4} \end{array} \right]}
\newcommand{\bracvec}[2]{\left[\begin{array}{c} {#1} \\ {#2} \end{array} \right]}

\title[Information Content of Slabs]{The Information Content of Projected Galaxy Fields}

\author[L. Porth et al.]{
Lucas Porth \orcidlink{0000-0003-1176-6346}$^{1,2,3}$\thanks{E-mail: lporth@uni-bonn.de},
Gary M. Bernstein \orcidlink{0000-0002-8613-8259}$^{2}$\thanks{E-mail:garyb@physics.upenn.edu},
Robert E. Smith \orcidlink{0000-0001-9989-2149}$^{1}$,
Abigail J. Lee \orcidlink{0000-0002-5865-0220}$^{2,4}$
\\
$^{1}$Astronomy Centre, Department of Physics \& Astronomy, University of Sussex, Brighton, BN1 9RH, UK\\
$^{2}$Department of Physics \& Astronomy, University of Pennsylvania,
Philadelphia, PA 19104, USA\\
$^{3}$Argelander-Institut f\"ur Astronomie, Universit\"at Bonn, Auf dem H\"ugel 71, 53121 Bonn, Germany\\
$^{4}$Department of Astronomy \& Astrophysics, University of Chicago, 5640 South Ellis Avenue, Chicago, IL 60637\\
}

\date{Accepted XXX. Received YYY; in original form ZZZ}

\pubyear{2021}

\begin{document}
\label{firstpage}
\pagerange{\pageref{firstpage}--\pageref{lastpage}}
\maketitle

\begin{abstract}
The power spectrum of the nonlinearly evolved large-scale mass
distribution recovers only a minority of the information
available on the mass fluctuation amplitude.  We investigate the
recovery of this information in 2D ``slabs'' of the mass distribution
averaged over $\approx100$~$h^{-1}$Mpc along the line of sight, as might be
obtained from photometric redshift surveys.  
We demonstrate a
Hamiltonian Monte Carlo (HMC) method to reconstruct the non-Gaussian
mass distribution in slabs, under the assumption that the projected
field is a point-transformed Gaussian random field, Poisson-sampled
by galaxies.  When applied to the \textit{Quijote} $N$-body suite at $z=0.5$ and
at a transverse resolution of 2~$h^{-1}$Mpc, the method recovers $\sim 30$ times
more information than the 2D power spectrum in the well-sampled limit,
recovering the Gaussian limit on information. At a more realistic
galaxy sampling density of $0.01$~$h^3$Mpc$^{-3}$, shot 
noise reduces the information gain to a factor of five improvement
over the power spectrum at resolutions of 4~$h^{-1}$Mpc or smaller.

\end{abstract}




\section{Introduction}
The potential constraining power of cosmological large-scale-structure surveys depends, at root, on the observed area and on the survey depth. As a theoretical description of the nonlinear density field is lacking, the full information cannot be retrieved and instead one performs cosmological analyses with the help of summary statistics. The most widely used statistic is the power spectrum, which captures the available information only for Gaussian fields. In the context of 3D fields \citep{Neyrincketal2009,Neyrincketal2011,Simpsonetal2013} and the cosmic convergence field \citep{Joachimietal2011,Seoetal2011,Simpsonetal2016,Giblinetal2018} linearization and clipping methods have been proposed that successfully remap some of the nonlinear information into the second order statistics. However the information gain can be strongly depleted once shot noise is taken into account. A more complete way to extract cosmological information from spectroscopic surveys can be formulated in terms of forward modelling approaches that do not need to compress the observed data. In particular the proposed models in \cite{KitauraEnsslin2008,JascheKitaura2010,JascheWandelt2013,Wangetal2014} and \cite{Jascheetal2015} have been successfully applied to N-body simulations and real galaxy redshift surveys \citep{LavauxJasche2016,Leclercqetal2017} to reveal a wealth of information of the dark matter density field and its phase space distribution in our nearby Universe. Due to the high dimensionality of the resulting posteriors, those models have mainly been used to reconstruct 3D cosmic fields when using a fixed cosmology. Recently the joint sampling of cosmology and fields has gained more traction \citep{LeclercqHeavens2021,Porqueresetal2021}. 

In this work we aim to test how well an idealized forward model performs for \emph{projected} tracer fields with a line-of-sight resolution of $\Delta \chi \approx100$~$h^{-1}$Mpc, such as might be acquired from imaging surveys using photometric redshifts. To assess the information contained in the resulting posterior we do not keep the cosmology fixed and allow the amplitude of the transformed power spectrum to vary. The information content can then be rephrased as the signal-to-noise (squared) of the estimated amplitude parameter after having marginalized over the projected mass field itself.

This paper is organised as follows: In \Sect{sec:Model} we introduce the hierarchical model adopted for our reconstruction and test the validity of our parametrizations on the Quijote simulation suite \citep{Quijote2020}. In \Sect{sec:Sampling} we give an overview of the Hamiltonian Monte Carlo sampling algorithm and lay out some specific choices we made for our implementation. In \Sect{sec:Results} we first validate our model on a suite of lognormal simulations, then apply it to the Quijote suite and finally compare the reconstruction confidence intervals of the hierarchical model to the expected signal-to-noise using standard 2-point analysis methods. In \Sect{sec:Conclusions} we summarize our findings, conclude and discuss future work.
\section{Model}\label{sec:Model}
For this work we suppose that we are given a set of galaxies on a two dimensional plane. We then cover the plane with a regular grid and assign the galaxies to the grid cells, yielding a set of galaxy counts $N\equiv\{N_c\}$. A hierarchical model connecting the observations to a Gaussian mass field $\deltalin$ with a (projected) power spectrum\footnote{Throughout this work we will refer to the volume normalised, projected power spectrum as the power spectrum.} $P$ parametrized by a set of parameters $\Pi_P$ can then be schematically written as:
\begin{align}
    &\condprob{\deltalin, \Pi_P, \Pi_B, \Pi_G}{N}
    \nonumber\\ &\hspace*{1cm}\propto
    \int \mathrm{d}\deltanl\; \condprob{N}{\deltanl,\Pi_B}
    \condprob{\deltanl}{\deltalin, \Pi_G} \nonumber \\
     & \hspace*{1cm}\phantom{\propto} \times
    \condprob{\deltalin}{\Pi_P}
       \mathbb{P}(\Pi_P) \mathbb{P}(\Pi_B,\Pi_G)\ ,
       \label{eq:4_Posterior}
\end{align}
where $\deltanl$ denotes a nonlinear mass field for which the galaxies are assumed to be biased tracers and $G$ is a gaussianization operator with parameters $\Pi_G$ that maps $\deltanl$ to $\deltalin.$ The model for galaxy bias is encoded in a bias operator $B$ with parameters $\Pi_B$. In the following we will give our modelling choices for each contribution to \Eqn{eq:4_Posterior}.
\subsubsection*{The Poissonian likelihood}
The first term models the way on how the set of discrete tracers is sampled onto the dark matter fluid, we assume this to follow an inhomogeneous Poisson process \citep{Layzer1956,Peebles1980}:
\begin{align}
    \condprob{N}{\deltanl,B,\bar{n}}
    &= \mathbb{P}\left(\bar{n}\right)
    \prod_c \mathrm{e}^{-n_c V} \frac{(n_c V)^{N_c}}{N_c!} \ ,
    \label{eq:4_PoissonLikeLong} \\
    n_c &\equiv \bar{n}\left(1+B\left[\deltanl\right]_c\right)
    \ , \label{eq:4_GalOverdensityLong}
\end{align}
where the intensity is given by the galaxy density field $n$ which depends on the mean number density of observed galaxies $\bar{n}$ and on the physical connection $B$ between the galaxies and dark matter. The product of the intensity with the underlying voxel volume $V$ then gives the expected number of galaxies in the corresponding pixel. Furthermore, we put a logarithmic prior on $\bar{n}$, such that one can perform an analytic marginalization:
\begin{align}\label{eq:4_PoissonLike}
    \condprob{N}{\deltanl,B}
    \nonumber \\
    &\hspace*{-1cm}\propto
    \int \frac{\mathrm{d}\bar{n}}{\bar{n}} \; \prod_c \mathrm{e}^{-n_c V} \frac{(n_c V)^{N_c}}{N_c!}
    \nonumber \\
    &\hspace*{-1cm}\propto
    \prod_c \left(1+B\left[\deltanl\right]_c\right)^{N_c}
    \int \mathrm{d}\bar{n}\; \bar{n}^{N_\mathrm{tot}-1}\mathrm{e}^{-\bar{n}V\sum_c\left(1+B\left[\deltanl\right]_c\right)}
    \nonumber \\
    &\hspace*{-1cm}\propto
    \left\{\sum_c \left(1+B\left[\deltanl\right]_c\right)\right\}^{-N_\mathrm{tot}} 
    \prod_c \left(1+B\left[\deltanl\right]_c\right)^{N_c} \ ,
\end{align}
where $N_{\mathrm{tot}}$ denotes the total number of observed tracers. With the galaxy counts $N$ as the only observables, there is a degeneracy between the gaussianization function $G$ and the bias functional $B$.  For this paper we will therefore assume the identity $B(\delta)=\delta,$ and we construct a ``galaxy'' catalog by Poisson sampling of the true mass distribution. In future work we will consider the combination of galaxy-count and weak-lensing observables, which will permit introduction of non-trivial bias, and a distinction between the mass and galaxy fields.

\subsubsection*{Gaussianization function}
The second term in \Eqn{eq:4_Posterior} corresponds to a physical model that evolves a linear density field to a nonlinear one. For this work we will assume a deterministic point transformation model of structure formation, such that
\begin{align}
    \condprob{\deltanl}{\deltalin, \Pi_G}
    =
    \prod_c \delta^D\left[ \deltanl_c - G^{-1}\left( \deltalin_c, \Pi_G \right) \right] \ ,
\end{align}
where $G$ is a function that aims at inverting structure formation. For this work we will consider two different forms of $G^{-1}$. The first is a logarithmic transformation, while the second one (which we dub DoubleLog) interpolates between two exponentials: 
\begin{align}
    &\mathrm{Logarithmic:}
    \nonumber \\
    &G^{-1}(\deltalin) = \mathrm{e}^{\deltalin - \sigma^2/2} - 1 \ ;
    \label{eq:4_LogTrafo}\\
    &\mathrm{DoubleLog:}
    \nonumber \\
    &G^{-1}(\deltalin) = n \ 
    \mathrm{e}^{a_1\deltalin - a_1^2\sigma^2/2}
      \left( 1+ \mathrm{e}^{(\deltalin-\delta_0)t} \right)^{\frac{a_2-a_1}{t}} - 1 
      \label{eq:4_Schechtertrafo} \\
    & \phantom{G^{-1}(\deltalin)} = n^\prime e^{\frac{\alpha_1+\alpha_2}{2} \deltalin}
      \cosh\left[ (\deltalin-\delta_0)t/2\right]^{\frac{a_2-a_1}{t}} - 1 
\end{align}
where $\sigma^2$ is the variance of the linear overdensity field and the normalization constants $n$ and $n^\prime$ are defined to yield $\langle\deltanl\rangle \equiv 0$. Note that the DoubleLog transformation is constructed to interpolate between two biased logarithmic transformations around a characteristic scale $\delta_0$ with a transition width described by $1/t$.  For the logarithmic transformation there are no free parameters $\Pi_G$ once the linear field's variance $\sigma^2$ is specified.  For the DoubleLog function, $\Pi_G=\{\alpha_1, \alpha_2, \delta_0, t\}.$ 
\subsubsection*{The Gaussian prior}
Assuming that the function $G$ completely gaussianizes the $\deltanl$, the linear field will be fully described by its correlation function $\xi$, which depends on $\Pi_P,$ such that we can write down a corresponding prior as
\begin{align}\label{eq:4_GaussianPriorRealSpace}
    \condprob{\deltalin}{\Pi_P}
    =
    \frac{1}{\sqrt{(2\pi)^{n_{\mathrm{pix}}}|\xi|}} 
    \exp\left[-\frac{1}{2}\sum_{c,c'} \deltalin_c \xi^{-1}_{cc'} \deltalin_{c'}\right] \ ,
\end{align}
where $n_{\mathrm{pix}}$ denotes the total number of pixels in the grid and $|\xi|$ is the determinant of the correlation matrix. In order to circumvent the computationally infeasible operations in this representation we evaluate the determinant and the convolution in \Eqn{eq:4_GaussianPriorRealSpace} in its harmonic basis indexed by a wavevector $k$:
\begin{align}\label{eq:4_GaussianPrior}
    \condprob{\deltalin}{P\left(\Pi_P\right)}
    \propto
    \left(\prod_k P_k^{-1/2}\right)
    \exp\left[-\sum_k \frac{ \left|\widetilde{\deltalin}_k \right|^2}{2P_k} \right]  .
\end{align}
For the remainder of this work we will assume that $P$ is parametrizable by a function that interpolates between two different power laws, with $\Pi_P=\{A,k_0,a_1,a_2,s\}$:
\begin{align}\label{eq:4_SchechterPk}
    P(k;\Pi_P)
    =
    A \left(\left(\frac{k}{k_0}\right)^{a_1s}+\left(\frac{k}{k_0}\right)^{a_2s}\right)^{-\frac{1}{s}} \ .
\end{align}
The parameters $k_0$ and $s$ determine the location and sharpness of the power-law transition. 

\begin{figure*}
    \centering
    \centering{
    \includegraphics[width=.98\columnwidth]{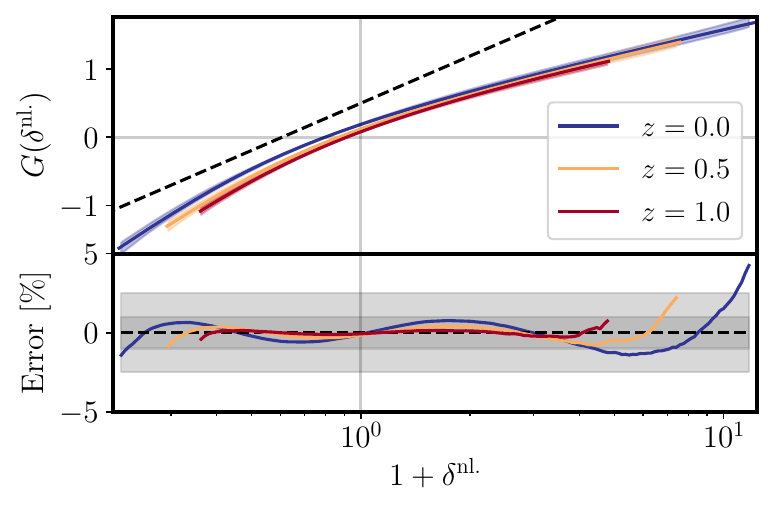}
    \includegraphics[width=.98\columnwidth]{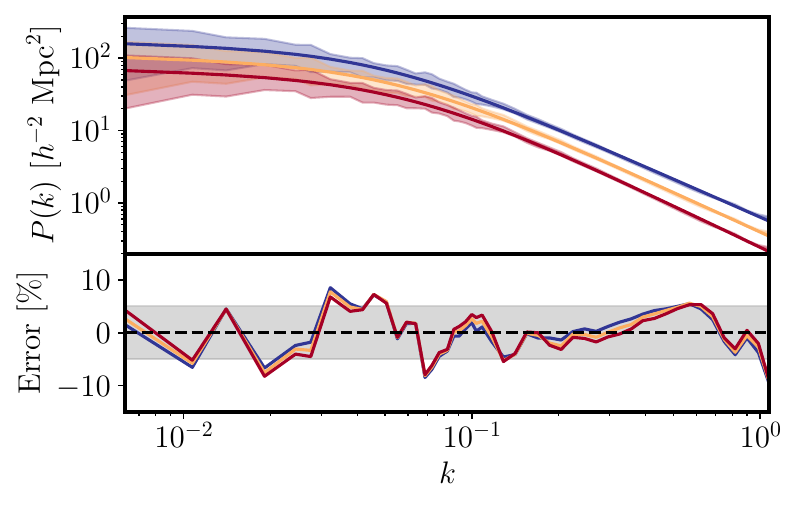}
    }
      \caption[Gaussianizing transformation accuracy in the Quijote suite]{{\emph{Left hand side}: Gaussianizing transformations for a transverse resolution of $ \approx 4$ $h^{-1}$Mpc for the three redshifts probed in this work. The shaded region in the upper panel corresponds to the standard deviation of the numerical transformation equations across the ensemble. The solid lines in the upper panel correspond to the best fit model \Eqn{eq:4_Schechtertrafo}. The black dashed line indicates the expected scaling from a logarithmic transformation model \Eqn{eq:4_LogTrafo}. The lower panel displays the relative error between the best fit model and the mean numerical transformation with the shaded regions displaying the $2\%$ and $1\%$ errorbands, respectively. \emph{Right hand side}: Same as the plot on the left, but for the power spectra of the transformed fields. In the lower panel we plot the $5\%$ errorband. In order to make the curves appear smooth the high-$k$ regime we switch to a logarithmic binning at these scales.}}
      \label{fig:MeasVsTheory_256}
\end{figure*}
\subsubsection*{Accuracy of parametrizations}
For assessing the applicability of the parametrizations \Eqn{eq:4_Schechtertrafo} and \Eqn{eq:4_SchechterPk} to N-body simulations we test their accuracy with help of the Quijote  suite. In particular, we make use of the  ensemble of $100$ high resolution simulations in which $1024^3$ particles were evolved within a $1h^{-1}\mathrm{Gpc}$ box. After retrieving the snapshots at $z\in\{1,0.5,0\}$ we assign the particles onto a regular mesh consisting of $256^3$ voxels using a NGP scheme. From those we create projected mass fields by specifying a projection depth and summing up the content in the corresponding voxels.  

For each mass slab $i$ we then employ inverse transform sampling to construct a linearising function $G^{\mathrm{num}}_i$ that maps the projected overdensity field to a field following a Gaussian distribution with zero mean and a variance matching the one we would have obtained when log-transforming the overdensity field. Averaging over all the $G^{\mathrm{num}}_i$ defines our numeric gaussianizing function $G^{\mathrm{num}}$ to which we fit the DoubleLog transformation \Eqn{eq:4_Schechtertrafo}. The best fit parameters then determine our model for $G^{-1}$. The results from this procedure for our chosen angular resolution of $\approx 4$ $h^{-1}$Mpc and a projection depth of $\approx 100 $ $h^{-1}$Mpc are shown in the left panel of \Fig{fig:MeasVsTheory_256} and we see that $G$ does give a percent-accurate fit for nearly all pixels. We furthermore note that while $G$ scales similarly to a logarithmic model around the mode of $\deltanl$, it does quite strongly deviate from such a model for moderately overdense and underdense regions. 

After having fixed the linearization procedure we transform each of the projected overdensity fields according to $G$ and compute the associated power spectra $P_i$. Again, we determine the best fit parameters of the model \Eqn{eq:4_SchechterPk} against the mean of the $P_i$ to define our final model for $P$. We show the numeric result in the right hand side of \Fig{fig:MeasVsTheory_256}. As for the transformations we find a reasonable agreement across all scales. Repeating the analysis described above for different transverse resolutions of approximately $16, \ 8 \ \mathrm{and} \ 2 \ h^{-1}$ Mpc, as well as for power spectrum fits to models where the logarithmic transformation \Eqn{eq:4_LogTrafo} had been applied we find that in all cases the chosen parametrizations provide a good enough fit for the main goal of this paper, i.e. to reasonably assess the information contained in the hierarchical model \Eqn{eq:4_Posterior}. We postpone a more thorough modelling of the power spectrum (i.e. by including BAO features or by directly linking it to differentiable Boltzmann codes) to future work.

\section{Sampling method}\label{sec:Sampling}
\subsection{Hamiltonian Monte Carlo Sampling}
We employ a Hamiltonian Monte Carlo (HMC) scheme \citep{Duaneetal1987} to efficiently sample from the high dimensional distribution in \Eqn{eq:4_Posterior}. This method evades the curse of dimensionality by exploring level sets of a distribution $\mathcal{P} \propto \mathrm{e}^{-\mathcal{H}}$ in which the Hamiltonian $\mathcal{H}$ is defined as 
\begin{align}\label{eq:4_HMCHamiltonian}
    \mathcal{H}(\vec{q}, \vec{p})
    &\equiv
    \frac{1}{2} \vec{p}^T\mathbf{M}^{-1}\vec{p} + \psi(\vec{q}) \ ;
    \\
    \psi(\vec{q}) &\equiv -\ln\mathbb{P}(\vec{q}) \ ,
\end{align}
where we assume the auxiliary momentum variables $\vec{p}$ to follow a Gaussian distribution, $\vec{p} \sim \mathcal{G} \left(0,\mathbf{M}\right)$.
From this formulation one can construct a valid Markov chain for the original posterior $\mathbb{P}$ by marginalizing over the momenta and for each drawn sample of $\vec{p}$ evolve the system to a new location $(\vec{q}',\vec{p}')$ in phase space according to the Hamilton equations of motion
\begin{align}
    \frac{\d \vec{q}}{\d t} 
    &= \frac{\partial \mathcal{H}}{\partial{\vec{p}}} 
    = \mathbf{M}^{-1}\vec{p}
    \nonumber \\
    \frac{\d \vec{p}}{\d t} 
    &= -\frac{\partial \mathcal{H}}{\partial{\vec{q}}}
    = - \nabla_{\vec{q}}\psi(\vec{q}) 
    \label{eq:4_HamiltonianEOM} \ . 
\end{align}
Due to numerical inaccuracies the Hamiltonian will not be exactly conserved along the trajectories. Thus, in order to still satisfy the detailed balance condition one then needs to invoke a Metropolis-Hastings rejection step prior to updating the chain with the value of $\vec{q}'$. For more complete reviews of HMC see e.g. \cite{Neal2011}.
\\\\
Adopting the HMC framework to the posterior \Eqn{eq:4_Posterior} we have $\vec{q} = \{\deltalin, \Pi_P\}$ and by making use of \Eqns{eq:4_GaussianPrior}{eq:4_PoissonLike} the potential $\psi$ becomes
\begin{align}\label{eq:4_Potential}
    \psi\left(\vec{q}\right)
    &= \psi_{\mathrm{Poiss}}\left(\deltalin\right) + \psi_{\mathrm{Gauss}}\left(\deltalin,\Pi_P\right)
    \ ; \\
    \psi_{\mathrm{Poiss}}
    &= N_{\mathrm{tot}} \log\left[ \sum_c \left( 1+G^{-1}\left(\deltalin_c\right) \right)\right]
    \nonumber \\
    &\hspace{.5cm}- \sum_c N_c \ln \left(1+G^{-1}\left(\deltalin_c\right)\right) \ ;
    \\
    \psi_{\mathrm{Gauss}}
    &= \frac{1}{2}\sum_k\left(\ln(P_k)+\frac{ \left|\widetilde{\deltalin}_k \right|^2}{P_k}\right) \ .
\end{align}
\subsection{Implementation specifics}
As for all sampling schemes there exist multiple knobs that need to be tweaked in order to facilitate an efficient exploration of our posterior. In this subsection we give a top-level overview of the choices for our implementation and refer the reader to \App{app:ExplicitExpressions} for more details.

To avoid performing a computationally infeasible number of $O(N_c^2)$ steps in the implementation of \Eqn{eq:4_HamiltonianEOM}, it is essential that the mass matrix $\mathbf{M}$ takes a sparse form in a readily accessible basis. For efficiency of the HMC chain, i.e. weakly correlated samples, $\mathbf{M}$ should 
approximate the Hessian of \Eqn{eq:4_Posterior}. Our most efficient solution is sparse in harmonic space and it allows for sampling of complex momenta from $\mathcal{G} \left(0,\mathbf{M}\right)$ at computational complexity bound by FFT operations at $O(N_c\log N_c)$, as well as for $O(N_c)$ complexity for the matrix vector product in \Eqn{eq:4_HamiltonianEOM}. We note that employing a diagonal mass matrix associated with the Hessian of the Gaussian part of the potential gave similarly good convergence properties.

As advocated by the standard literature we implement a leapfrog integrator to discretise the evolution equations \Eqn{eq:4_HamiltonianEOM}. This method is expected to be effective as it is a second order symplectic discretization scheme that will conserve the Hamiltonian for a well chosen step-size. We also check whether some versions of higher order symplectic integrators (see \cite{CreutzGocksch1989,Yoshida1990,McLachlan1995} for the original formulations, or \cite{Hernandezetal2021} for a first application to cosmology) result in an effective speedup; while for models using a fixed cosmology the leapfrog integrator remains the most efficient routine we find that for models with varying cosmology the fourth order integrator of \cite{McLachlan1995} yields the largest effective sample size per unit time. 

In order to choose a useful step-size for the integrator we apply a dual averaging scheme \citep{Nesterov2009,HoffmanGelman2014} during the burn-in stage that iteratively adapts the step-size to a value that will result in some specified acceptance rate $\delta$ during the sampling phase. Once burn-in is over we fix the step-size to its final value $\epsilon$ in the iteration. 

Due to the high dimensionality of the problem we also need to worry about the memory footprint of the chain outputs. In this work we are mainly concerned with the chains of the power spectrum parameters $\Pi_P$ and therefore we save those as a whole, but we save the latent field parameters $\deltalin$ of only a fraction of the pixels. As the convergence properties of the $\deltalin_c$ depend on the number of tracers in the corresponding pixel we make sure that our selection does include sufficiently many overdense and underdense regions. For assessing the convergence of the latent field in harmonic space we additionally store a representative selection of chains in this basis.
\section{Results}\label{sec:Results}
\subsection{General strategy}\label{ssec:Strategy}
For the remainder of this work we will solely concern ourselves with models $\mathcal{M}$ that vary the power spectrum amplitude $A$ jointly with the cosmological field $\deltalin$ and fix the remaining power spectrum parameters to their best-fit values. We furthermore introduce a nominal value $A^*$ for the power spectrum amplitude that is used to build the mass matrix. The nominal value is related to the best-fit value $A^\mathrm{f}$ of the amplitude as  $A^* = \beta A^\mathrm{f}$, where in our case we let $\beta \in [2/3,3/2]$. We can then take $A^*$ to be the initial value $A_0$ for the amplitude in the chain. A possible starting position of the latent field, $\deltalin_0$, can be chosen as a random Gaussian field constructed to match the true power spectrum with a strongly reduced amplitude. 

When running the model $\mathcal{M}$ with these initial conditions we found that the burn-in period becomes very prolonged as the starting point is in strong conflict with the Poisson likelihood and a very small step-size becomes necessary to navigate the chains to their stationary territory\footnote{We tried various other initialization choices for $\deltalin_0$ and $A_0$. For each of them we found the same pathological behaviour.}. To circumvent most of the complexity we adopt a nested burn-in strategy where in a first step we run a simpler model $\mathcal{M}'$ in which we also fix $A \equiv A^*$. Burning in this model with $\deltalin_0$ is fast and choosing some sample from $\mathcal{M}'$ once stationarity is reached yields a better starting configuration $\deltalin_0$ for the full model $\mathcal{M}$ that does now burn in much quicker.

Additionally, we note that our procedures for generating mock data, as well as the sampling procedure, are stochastic and are drawn according to some random seeds $r$ for each chain:
\begin{enumerate}
    \item The cosmic initial condition $r_{\mathrm{cosmo}}$ that gives rise to to `true` projected density field, i.e. which realization and spatial ``slab'' of the Quijote simulation are used;
    \item The Poisson sampling process $r_{\mathrm{Poiss}}$ that selects ``galaxies'' from the mass distribution;
    \item The $r_{\mathrm{ini}}$ used to initialize $\deltalin$ in the reconstruction algorithm;
    \item The $r_{\mathrm{mom}}$ for the Markov chain, i.e. the draws from the multivariate normal momentum distribution, and for the Metropolis rejection step.
\end{enumerate}

If we want to make a solid prediction about the information content in the power spectrum amplitude $A$ we would formally need to marginalize over a large set of seed configurations. We can get rid of one dimension when making the assumption that $r_{\mathrm{ini}}$ and $r_{\mathrm{mom}}$ do not influence each other, and we collect both seeds in a new one,  $r_{\mathrm{hmc}}$. Checking the dispersion of the outcomes when varying over $r_{\mathrm{hmc}}$ for fixed $r_{\mathrm{cosmo}}$ and $r_{\mathrm{Poiss}}$ is then equivalent to assessing the convergence property of the chains, i.e. by virtue of the Gelman-Rubin diagnostics \citep{GelmanRubin1992}. Varying over the remaining two seeds is necessary and we do this for our analysis.
\begin{figure*}
\centering
    \centering{
    \hspace{1.75cm}
    \includegraphics[width=1.4\columnwidth]{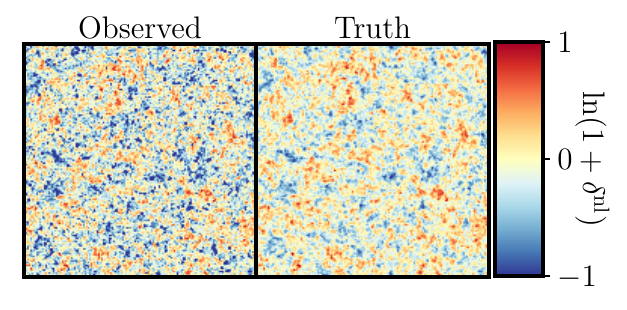}\\
    \includegraphics[width=1.04\columnwidth]{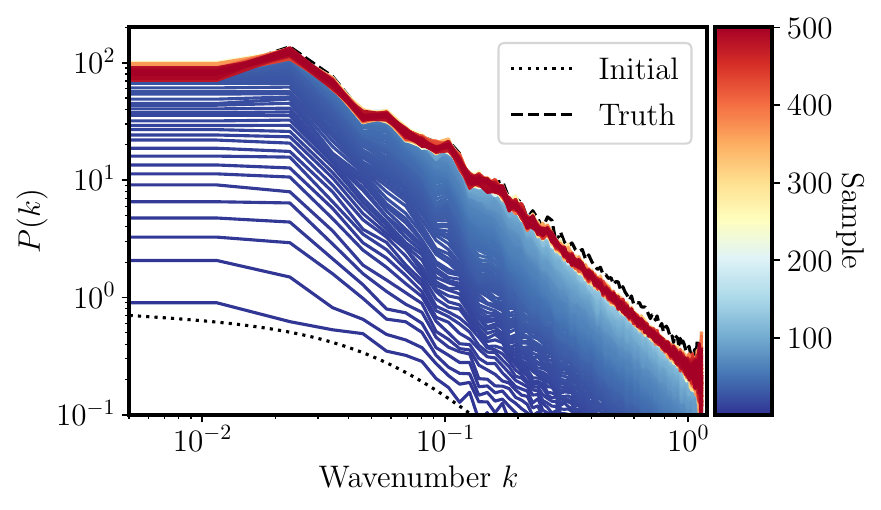}
    \includegraphics[width=.92\columnwidth]{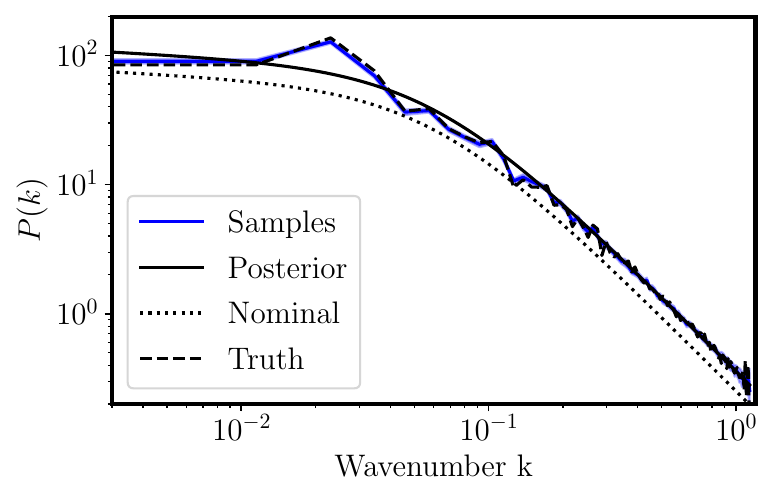}\\
    \includegraphics[width=1.4\columnwidth]{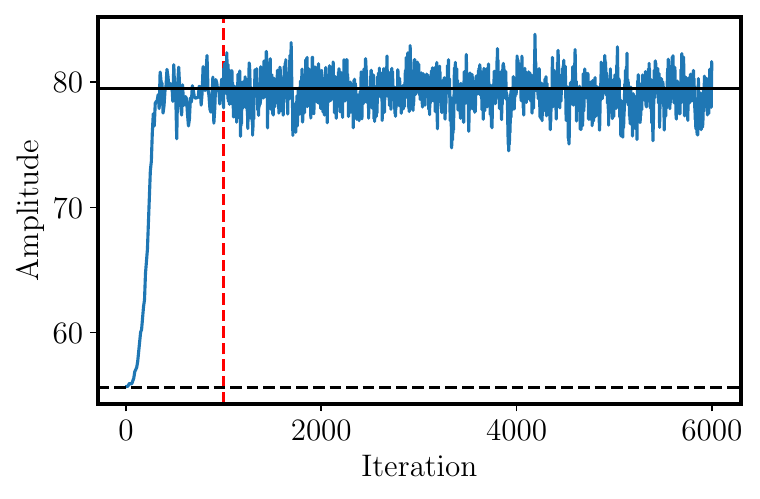}
    }
      \caption[HMC on Lognormal mock]{{Sampling of the posterior \Eqn{eq:4_Posterior} on a lognormal mock catalog. \emph{Upper panel:} Comparison of the observed galaxy field to the underlying true field. \emph{Middle left panel:} Power spectra of the proposed latent fields during burn-in of the simplified model $\mathcal{M}'$. \emph{Middle right panel:} Results of the sampling stage when using the full model $\mathcal{M}$. The blue errorband indicates the variance of the power spectra measured from the proposed latent fields whereas the black solid line gives evaluates the theoretical power spectrum model with the mean of the proposed values for the amplitude. \emph{Lower panel:} The amplitude chain of the model $\mathcal{M}$ during the burn-in stage (left of the red dashed line) and the sampling stage. The solid black line indicates the value $A^{\mathrm{f}}$ that was used for generating the mock data while the dashed black line indicates the nominal value $A^*=0.7A^\mathrm{f}$.}}
      \label{fig:LNMockSample}
\end{figure*}
\begin{table*}
\centering
\begin{tabular}{|c|c|c|c|c|c|c|c|}
\hline
Name & 
Nside & 
Depth \ \ [$\Delta_{\mathrm{pix}}$] &
$G^{-1}$ & 
Snapshots &
Poisson realizations &
Tracer densities &
Initial conditions \\ 
\hline
DoubleLog\_HighRes  & 512 & 50  & DoubleLog & 5   & 4 & 4 & 6  \\ \hline
DoubleLog\_BaseRes & 256 & 25  & DoubleLog & 100 & 4 & 4 & 8  \\ \hline
DoubleLog\_LowRes   & 128 & 12  & DoubleLog & 5   & 8 & 4 & 6  \\ \hline
DoubleLog\_vLowRes  & 64  & 6   & DoubleLog & 5   & 8 & 4 & 6  \\ \hline
Log\_HighRes & 512 & 50  &    Log    & 5   & 4 & 4 & 6  \\ \hline
Log\_BaseRes & 256 & 25  &    Log    & 5   & 4 & 4 & 6  \\ \hline \hline
DoubleLog\_HighRes\_mcl & 512 & 50  & DoubleLog & 10 & 4 & 5 & 6  \\ \hline
DoubleLog\_BaseRes\_mcl & 256 & 25  & DoubleLog & 10 & 4 & 5 & 6  \\ \hline
DoubleLog\_LowRes\_mcl & 128 & 12  & DoubleLog & 10 & 4 & 5 & 6  \\ \hline
DoubleLog\_vLowRes\_mcl & 64 & 6  & DoubleLog & 10 & 4 & 5 & 6  \\ \hline
Log\_BaseRes\_mcl & 256 & 25  & Log & 10 & 4 & 5 & 6  \\ \hline
DoubleLog\_BaseRes\_lf & 256 & 25  & DoubleLog & 10 & 4 & 5 & 6  \\ \hline
\end{tabular}
\caption{Parameter settings for the various ensemble runs at $z=0.5$ performed on snapshots from the Quijote simulation suite. For each of the chains the four tracer densities that are varied over are $0.002, \ 0.005, \ 0.01, \ \mathrm{and} \ 1.0$ tracers per inverse $h^{-3}$Mpc$^3$. The DoubleLog$\_$BaseRes run probes the whole Quijote ensemble and consists of the most $(100\times 4 \times 4 \times 8 = 12800)$ individual chains. Each chain consists of $5000$ Hamiltonian samples. The first six runs are set up according to the description in the main text while for the latter six we use a more restrictive value of $\delta$ in the dual averaging algorithm. The last of those is performed using the standard leapfrog integrator while for all the other runs we use the fourth order symplectic integrator of \citet{McLachlan1995}. We do furthermore repeat the DoubleLog$\_\textrm{<resolution>}\_$mcl runs for redshifts of $z \in \{0,1\}$.}\label{tab:EnsembleRuns}
\end{table*}
\subsection{Validation on lognormal simulations}
While the point transformations $G$ do a good job in removing nonlinearities of the mass field, they are not sufficient to fully gaussianize the field, which renders the prior \Eqn{eq:4_GaussianPrior} formally incorrect. In order to test our implementation we apply it to an ensemble of mock catalogs drawn from truly lognormal mass distributions. In particular, we obtain a tracer realization as follows:
\begin{enumerate}
    \item We specify a resolution and spatial extent of the slab, as well as a tracer sampling density $\bar{n}$;
    \item We generate a 2d Gaussian random field $g$ having a power spectrum $P_g$ that is tuned to match the best-fit power spectrum of the Quijote ensemble of the corresponding slab specifics;
    \item We generate a lognormal field as $\delta^{\mathrm{ln}}$ by applying the transformation \Eqn{eq:4_LogTrafo} to $g$;
    \item We Poisson-sample tracer ``galaxies'' into the pixels
\end{enumerate}
As an example, we show in \Fig{fig:LNMockSample} the results of a single chain run on a $256^2$ grid. Here, we chose to evolve the system for $40$ time steps before updating the chain\footnote{In order ensure detailed balance and to avoid resonant behaviour we first decide whether to integrate forward or backward in time and then uniformly pick a sample from the resulting trajectory. We postpone the investiagtion of more thorough schemes (see i.e. \cite{Betancourt2016} or Appendix A of \cite{Betancourt2017}) that take into account the numerical inaccuracies of the symplectic integrators when selecting the trajectory to future work.}. Furthermore, we set $\beta \equiv 0.7$ and choose $\bar{n}\equiv0.005 \ h^3\mathrm{Mpc}^{-3}$. Looking at the first burn-in stage, using the model $\mathcal{M}'$, we see that the latent field has burned in to the nominal power spectrum after around $100$ iterations. Moving to the full model $\mathcal{M}$ we see that two measures of the power spectrum---the spectra of the latent $\deltalin$ fields of the samples, and the values of amplitude $A$ at each sample---give results that are consistent with each other and with the true mass field.  During the burn-in stage of $\mathcal{M}$, the chain of the amplitude $A$ evolves from the nominal value to the true amplitude that was used for generating the mock data\footnote{We note that for this example we use a less efficient step-size adaptation algorithm to explicitly showcase the evolution of the amplitude from the prior to the truth.}. After burn-in has finished the chain oscillates around the true value $A^{\mathrm{f}}$ and provides an unbiased estimate. 
\subsection{Application to the Quijote simulation suite}
\begin{figure*}
    \centering
    \centering{
    \includegraphics[width=\linewidth]{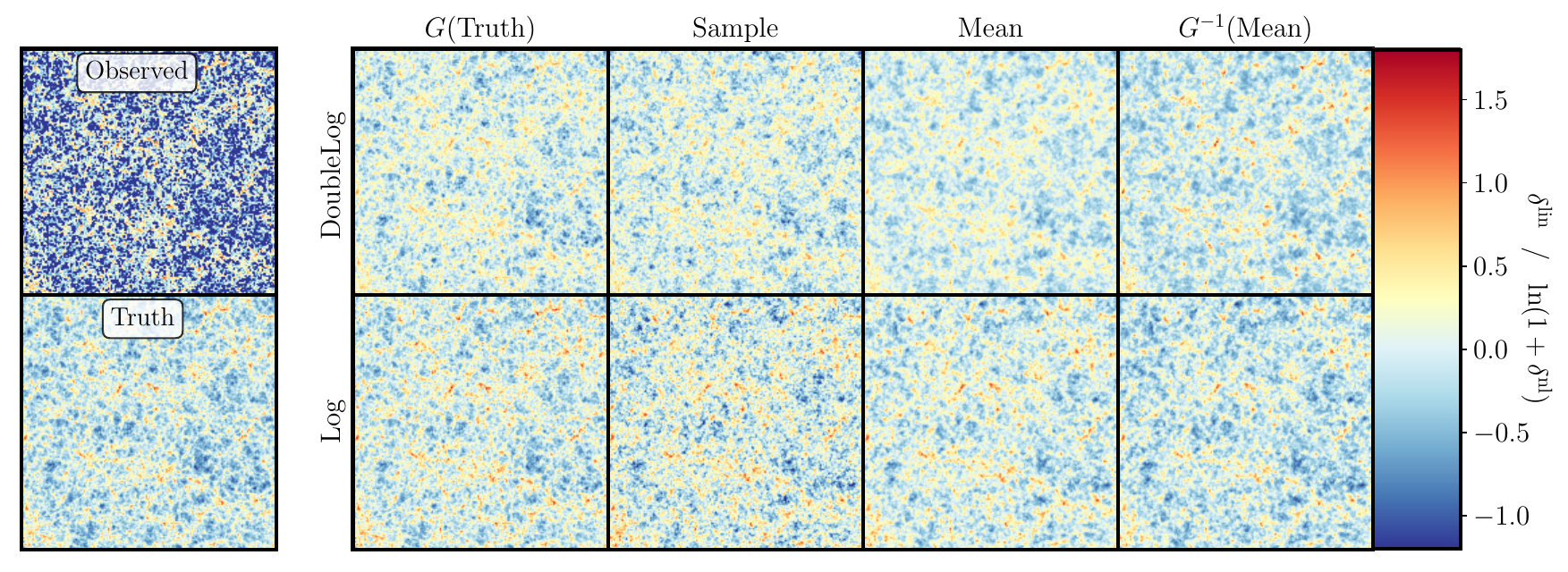}}
      \caption[Latent field reconstruction on one Quijote mock]{Latent field reconstruction on a Quijote slab using the base resolution. The leftmost plots show the observed tracer counts (top) and the underlying true mass field (bottom).
      In remaining panels show various measurements obtained from running the reconstruction using the DoubleLog transformation (top) or the lognormal transformation (bottom). In particular, the columns display the linearized true mass field (first), a sample from the chain (second), and the mean of all samples for the linear (third) and nonlinear (fourth) fields. For both runs we chose $\bar{n}=0.002 h^3\mathrm{Mpc}^{-3}$ and matched all the random seeds. }
      \label{fig:QuijoteRecLin}
\end{figure*}
\subsubsection*{Overview of ensemble runs}
We now turn to the runs on the Quijote ensemble. As discussed in \Sect{ssec:Strategy}, we vary cosmological ensembles, as well as Poisson sampling realizations. Additionally, we aim to investigate how the reconstruction confidence of our model is affected by the pixel resolution and the sampling density of tracers. In \Tab{tab:EnsembleRuns} we summarize the configuration details for each analysis used in this work.

For each chain we adopt the nested burn-in strategy: we sample from model $\mathcal{M}'$ for 500 times before switching to $\mathcal{M}$ which we burn in for another $1000$ steps after which the $5000$ samples that will be used for the subsequent analysis are generated. In order to obtain a new sample we evolve the equations of motion \Eqn{eq:4_HamiltonianEOM} for $40$ (jittered) time steps. To test the convergence of the chains we furthermore save the full chains for around $5-20$ per cent of the latent field pixels, depending on the grid resolution.
\begin{figure*}
    \centering
    \centering{
    \includegraphics[width=.98\columnwidth]{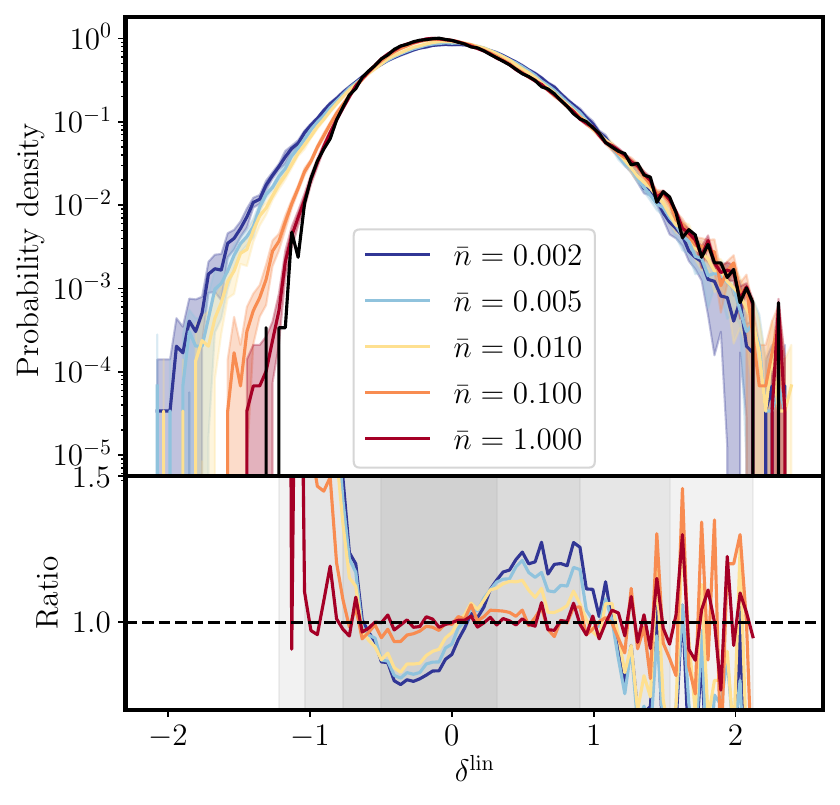}
    \includegraphics[width=.98\columnwidth]{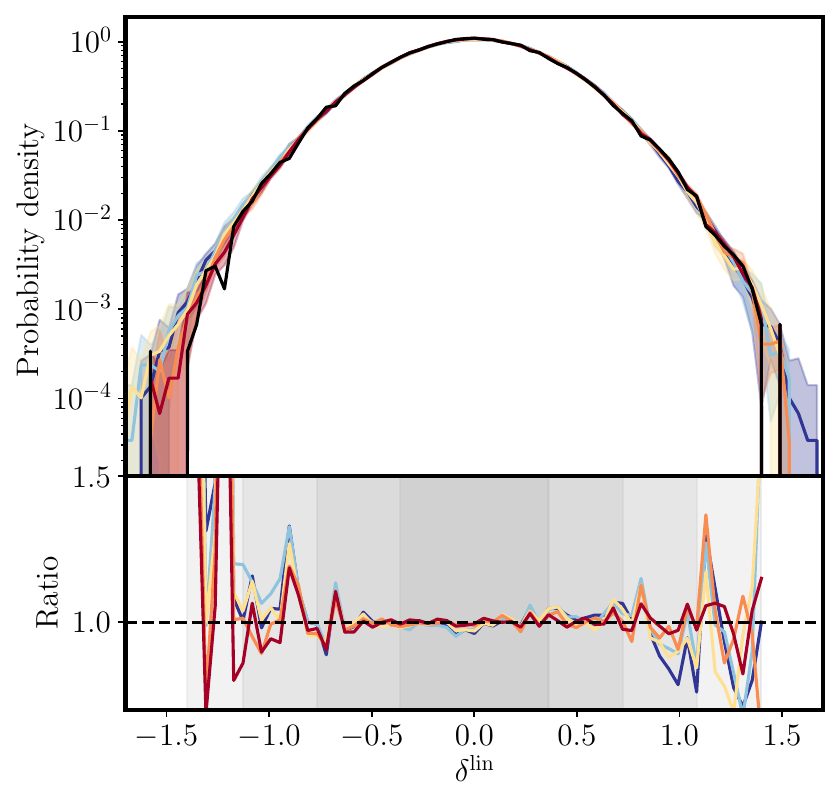}
    }
      \caption{Reconstruction accuracy of the one point pdf of the linear density field when using the logarithmic (left) or DoubleLog (right) transformation equation. The upper panels show the linearized true mass field (black solid line), as well as the range of latent field one point distribution that are predicted by our forward model when using different sampling densities of tracers (colored lines $+$ error bands). The bottom panels show the ratio between the true and the reconstructed pdfs and the vertical grey bands display the intervals in which 68, 95, 99.7 and 99.95 per cent of the true data resides.}
      \label{fig:RecPdf}
\end{figure*}
\subsubsection*{Reconstruction accuracy}
Before moving to the results of the ensemble we show in  \Fig{fig:QuijoteRecLin} the real-space latent field reconstructions of a single chain when using the DoubleLog transformation \Eqn{eq:4_Schechtertrafo} or the logormal one, \Eqn{eq:4_LogTrafo}. We see that, as expected, neither point transformation removes the filamentary structure, and the latent $\deltalin$ are not Gaussian fields.  The DoubleLog model does, however, produce a more slightly more Gaussian PDF, with a less extended tail of overdensities as compared to the lognormal model (see \Fig{fig:RecPdf}).  

We can predict this feature from \Fig{fig:MeasVsTheory_256} in which the gaussianizing transformation grows more slowly at high density than the logarithmic curve and thus will map a larger range of overdensities to an equal sized bin in the transformed field. Since, however, the high-density tail of tracers is best observed and constrained by galaxy counts, the logarithmic model (incorrectly) extrapolates this tail into the underdense regime as enforced by the Gaussian part of the potential \Eqn{eq:4_Potential}. We quantify the $\bar{n}$ dependence of this effect in \Fig{fig:RecPdf} and find that one would need an unrealistically high tracer density in order to faithfully reconstruct the one-point statistics of the underlying mass field using the Lognormal function. We can also verify in \Fig{fig:RecPdf} that the DoubleLog model does not suffer from this effect and therefore produces a faithful gaussianization at the one point level, independent of $\bar{n}$.
Turning back to \Fig{fig:QuijoteRecLin}, we see that on the level of an individual sample both models are confident in sampling similar structures in overdense regions while for underdense spots the models resort to their gaussian prior and will therefore not be able to predict the filamentary structure by themselves. Finally, for the mean field and its nonlinearized version we again observe a good reconstruction of the high ends of the density field while regions with little data information appear washed out.

\begin{figure}
    \centering
    \centering{
    \includegraphics[width=1\columnwidth]{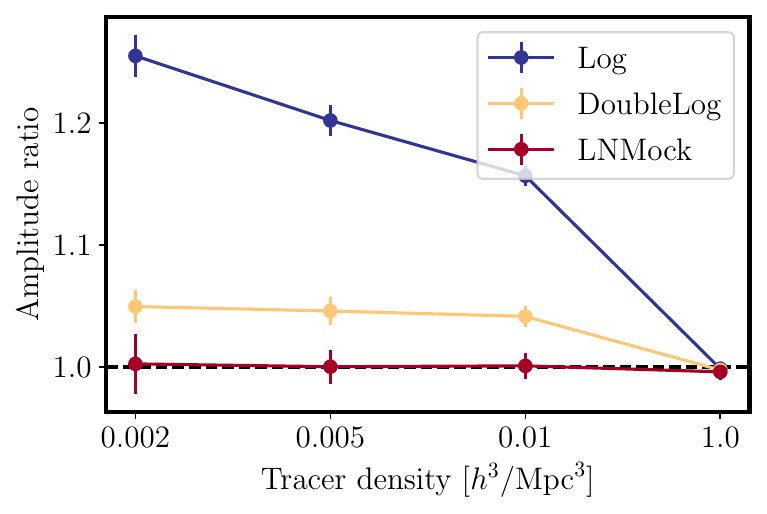}
    }
      \caption[Reconstructed power spectrum amplitude ratio]{{Ratio of the reconstructed power spectrum amplitudes as compared to the best-fit value $A^*$ across the ensemble for different tracer overdensities.}}
      \label{fig:AmplitudeBias}
\end{figure}

Due to the (non-Gaussian) filaments in the latent field, we might find that reconstruction with the Gaussian prior of \Eqn{eq:4_GaussianPrior}  yields a slightly biased amplitude. In \Fig{fig:AmplitudeBias} we quantify the magnitude of this effect and its dependence on the galaxy sampling density on the DoubleLog$\_$BaseRes and Log$\_$BaseRes runs. As a sanity check we also include the lognormal reconstructions LNMock$\_$BaseRes and we find that they give unbiased results, as should occur when the probability being used by the HMC is precisely that from which the data are drawn. We also see that when we assume that the $N$-body fields are point-transformed Gaussian fields, the DoubleLog transformation produces a much smaller bias on $A$ than the lognormal one. We again attribute this to the DoubleLog transform yielding a better gaussianization of the nonlinear field at the one-point level.  The extendend underdensity tail that the logarithmic model predicts can only be matched by over-estimating the power spectrum amplitude. The fewer tracers are observed, the more severe this effect and the resulting bias does become. As the DoubleLog model does not suffer from this feature its bias remains nearly constant across all $\bar{n}$. Repeating the analysis for different gridding scales we find that the bias of the DoubleLog model grows slightly with improving resolution while the logarithmic model is more strongly affected.  For both transformations, we find that increasing tracer density lowers the amplitude bias, as the tracers provide stronger constraints on the true mass field and the choice of generative model for $\deltanl$ is less important.
\begin{figure*}
    \centering
    \begin{subfigure}{1.16\columnwidth}
    \vspace{.5cm}
    \subfloat{\includegraphics[width=1\columnwidth]{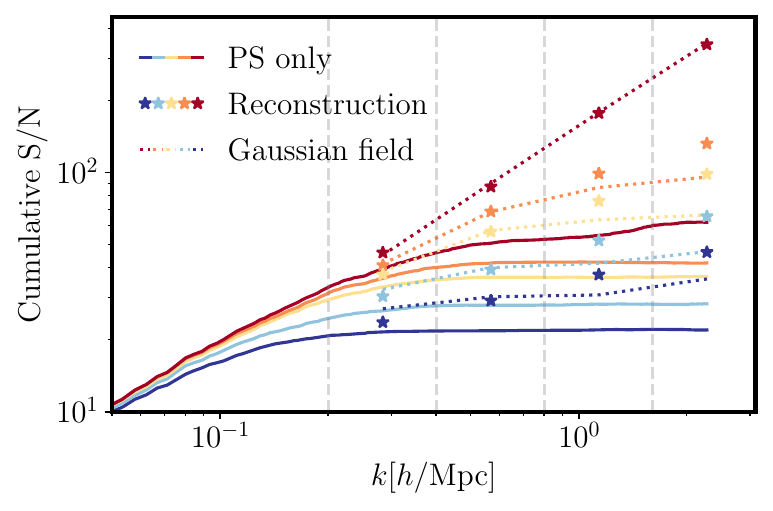}}
    \end{subfigure}
    \begin{subfigure}{.82\columnwidth}
    \vspace{-.5cm}
    \subfloat{\includegraphics[width=1\columnwidth]{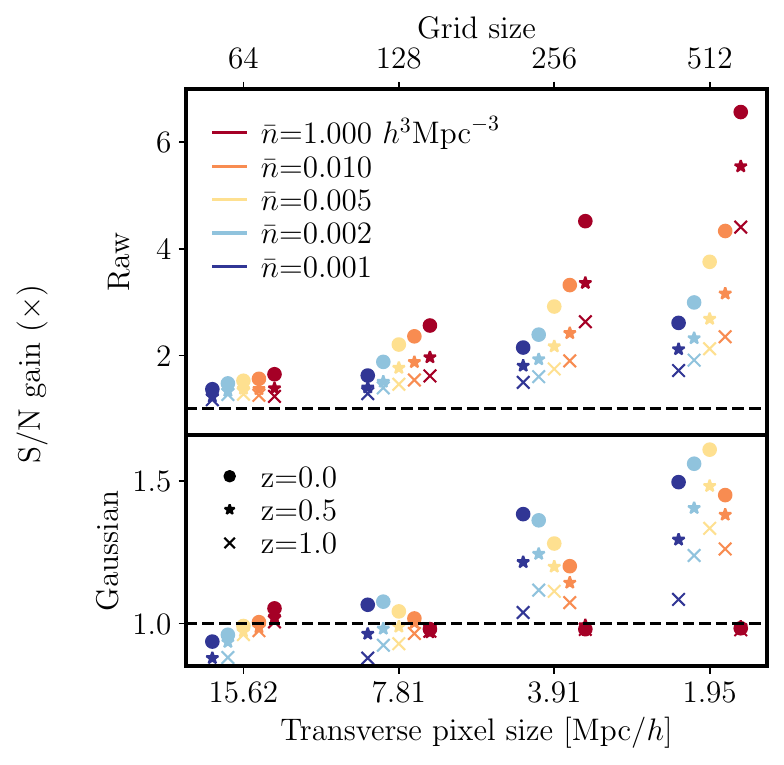}}
    \end{subfigure}
      \caption[Cumulative signal-to-noise for HMC vs baseline analysis]{{\emph{Left hand side}: Cumulative signal-to-noise ratio on the linear power spectrum amplitude $A$ at $z=0.5$ when using the power spectrum} (solid lines), a Gaussian field with a shot noise contribution matched to the DoubleLog transformation (dashed lines) or the hierarchical model \Eqn{eq:4_Posterior} with the DoubleLog transformation (star markers). The five different colors indicate the sampling densities of the tracers (see legend on the right hand side plot) and the different sets of points correspond to the results when running the reconstruction method on different resolutions, see \Tab{tab:EnsembleRuns} for the details. The grey dashed lines indicate the Nyquist frequencies for each of the probed grid resolutions. \emph{Right hand side}: Signal-to-noise gain of the reconstruction method with respect to the standard analysis in terms of the raw power spectrum (top) and the Gaussian field (bottom) for the three redshifts probed in this work.}
      \label{fig:SNIncrease}
\end{figure*}

\subsubsection*{Information content}
To assess on how much additional information the hierarchical model \Eqn{eq:4_Posterior} contains in comparison to a traditional 2-point analysis, we compare the reconstruction confidence of the power spectrum amplitude chain $\mathcal{C}_A$ to the expected variance of the measured power spectra. The first quantity can simply be determined by averaging the signal-to-noise of the $\mathcal{C}_A$ over the ensemble,
\begin{align} \label{eq:4_SNhier}
    \left(\frac{S}{N}\right)_{\mathrm{hier.}}
    \equiv
    \left\langle\frac{\mathbb{E}(\mathcal{C}_{A})}{\sigma(\mathcal{C}_{A})}\right\rangle_{\mathrm{chains}} \ .
\end{align}
For obtaining the corresponding measure from the standard analysis we follow the procedure put forward in \cite{RimesHamilton2005, LeePen2008} and \cite{ Neyrincketal2011} and estimate the cumulative information content in the projected power spectra of the untransformed field $P$ about the shot noise subtracted spectra $P^{-s}$,
\begin{align} \label{eq:4_SNtrad}
    \left(\frac{S}{N}\right)^2
    (< k)\equiv
    \sum_{i,j\in\mathcal{R}_k} 
    \left(\frac{P^{-s}_i}{P^{-s}_i+P^s_i}\right)
    \mathrm{r}^{-1}_{ij}
    \left(\frac{P^{-s}_j}{P^{-s}_i+P^s_j}\right) \ ,
\end{align}
where $P^s$ denotes the power spectrum induced by the shot noise, $r$ stands for the normalised correlation matrix of $P^{-s}$ and the sum runs over all index pairs for which $k_{i}, k_j<k$. For a truly Gaussian field with negligible shot noise, the cumulative information is simply the number of 2d Fourier modes available. To minimize the noise when estimating $r$ and therefore get a stable inverse, we obtain this quantity from $2000$ Quijote simulations. Those were run on the standard resolution, i.e. using $512^3$ dark matter particles and we construct the slabs and tracer counts as described in \Sect{sec:Model}. We include the Hartlap factor \citep{Hartlapetal2007} in the computation of the inverse correlation matrix to render it sufficiently unbiased\footnote{For assessing the impact of the mass resolution of the underlying $N$-body simulations we repeat the analysis described below using $3000$ nearly independent slabs of the $100$ Quijote HR simulations that were used for the Hamiltonian reconstruction. The results are fully consistent with the ones presented below.}. 

In the left hand side of \Fig{fig:SNIncrease} we compare the two measures \Eqns{eq:4_SNhier}{eq:4_SNtrad} for the DoubleLog models of different resolutions at a redshift of $z=0.5$. Firstly, we see that that the cumulative signal-to-noise of the traditional analysis starts to diverge from the Gaussian scaling at around $k=0.1 h\mathrm{Mpc}^{-1}$ and quickly plateaus thereafter. In contrast, for the hierarchical model the confidence on the linear amplitude has a significant dependence on the tracer density and resolution and does always yield more information than the power spectrum based method. For a high-sampling limit of $\bar{n} = 1 h^3\mathrm{Mpc}^{-3}$ \Eqn{eq:4_Posterior} recovers the ideal Gaussian information (mode count) for scales as small as $2 h^{-1}\mathrm{Mpc}$\footnote{In \App{app:CorrelationMatrix} we motivate these results from the correlation structure of the spectra of the underlying fields}. In the upper panel on the right we quantify the information ratio of both analysis methods for $z \in \{0.0,0.5,1.0\}$. We see that in addition to increasing the tracer density, additional information can be recovered when reducing the transverse scale of the reconstruction or when using mass fields at a later stage of the cosmic evolution. For example, at $z=0.5$ the information is at least quadrupled at translinear scales for an attainable shot noise level of $\bar{n} = 0.01 h^3\mathrm{Mpc}^{-3}$. 
In the left hand side of \Fig{fig:SNIncrease} we also show the information of the hierarchical model and the information contained in a discretely sampled Gaussian field, 
\begin{align}
    \left(\frac{S}{N}\right)^2
    (< k)\equiv
    \sum_{i}\frac{N_i}{2}\left(\frac{\bar{n}_{\mathrm{eff},i}P_i}{1+\bar{n}_{\mathrm{eff},i}P_i}\right)^2 ,
\end{align}
where $N_i$ denotes the number of modes in the corresponding $k$-bin and
for which the (scale dependent) effective sampling density $\bar{n}_{\mathrm{eff}}$ is taken to be the inverse of the shot noise spectrum of the transformed field. 

In the bottom right panel of \Fig{fig:SNIncrease} we show the information ratio of the hierarchical model and the discretely sampled Gaussian field. Firstly, we explicitly see that for the highest probed sampling density of tracers the hierarchical model contains the same information as the Gaussian field, which in that case also asymptotes to the mode count limit. Secondly, for the other tracer densities we see that the information ratio appears to increase with improved resolution and it starts being significantly above unity for transverse resolutions of less than $4 h^{-1}\mathrm{Mpc}$, signifying that when using a field based likelihood one is able to re-capture information that would have otherwise been ``lost''` in the shot noise. However, we also see that for a very low resolution the hierarchical model seems to contain less information than the Gaussian field. We attribute this to two facts, namely to the slight incorrectness of the point transformation model, as well as to the marginalization over $\bar{n}$ that is implicitly included in \Eqn{eq:4_Posterior}. 

\section{Conclusions}\label{sec:Conclusions}
In this paper we explored the information content of 2d ``slabs''
of the evolved mass distribution of the universe, when analysed by
sampling realizations of this projected mass field from a
simple hierarchical model conditioned on the projected galaxy field.
The model, described in \Sect{sec:Model}, was composed of a Poisson
likelihood for the discrete galaxy distribution and a Gaussian prior
on the linearized density field. 
\Sect{sec:Sampling} introduces the HMC sampling algorithm; our
implementation choices---namely a sparse non-diagonal mass matrix, a symplectic fourth order integration scheme, and automatic step-size
adaptation---yield a feasible and convergent chain for 2~$h^{-1}$Mpc
resolution of a 1~$h^{-1}$Gpc square slab.

In \Sect{sec:Results} we showed that the HMC successfully recovers the
power spectrum amplitude of lognormal mock catalogs, for which the
HMC has the exactly correct probability.  Moving to the Quijote
$N$-body simulation suite at $z=0.5$, for which a linearized Gaussian model is
incomplete, we found that there was some bias in the reconstructions, but
the DoubleLog function which very nearly Gaussianizes the
one-point distribution also recovered the linearized power spectrum
with few-percent bias.  With high galaxy density (low shot noise), the
point-transform reconstruction recovered nearly all of the information
on the amplitude, i.e. the Gaussian limit where information equals the
mode count, as highlighted in Figures~\ref{fig:SNIncrease} down to
resolutions of 2 $h^{-1}$Mpc, whereas the power spectrum captured only the
information for scales larger than $\approx20$~$h^{-1}$Mpc, after which it saturates.

In the presence of shot noise, the information in the HMC
reconstruction is of course degraded, but remains 4--5$\times$ higher
than for the power spectrum for sampling densities of $\bar n =
0.01$~$h^3$Mpc$^{-3}.$

This finding suggests that linearization methods proposed for improved
information retrieval in spectroscopic galaxy surveys can be applied
with substantial benefit to photometric galaxy surveys. This HMC
field-sampling method can be extended to multiple line-of-sight slabs,
and it will be straightforward to add the probability of weak lensing
shear or convergence observations to the hierarchical model.  Thus in
future work we will develop techniques for replacing the
``$3\times2$-point'' data vectors that are now the standard for
cosmological analyses of lensing$+$galaxy surveys with field sampling
that yields much more precision from the same data. 

Before closing we recall that the model presented in this work was intended to be idealized such that the computational cost for getting reasonable estimates of the information recovered by our field based reconstruction method was feasible. For obtaining robust cosmological constraints one needs to refine some modelling choices; i.e. include a different (yet still deterministic) model for structure formation, a power spectrum model obtained from a differentiable Boltzmann solver, and potentially also some modifications to the Poisson assumption for the sampling of tracers \citep{Nguyenetal2021}. We expect that the information content of such a model will still be comparable to the results found in this work and therefore yield more information than a traditional analysis.
%
%

\section*{Acknowledgements}
We thank the anonymous referee for helpful comments. LP acknowledges support from a STFC Research Training Grant (grant
number ST/R505146/1) and from the DLR grant 50QE2002. 
GMB acknowledges support for this work from US Department of Energy
grant DE-SC00079014 and National Science Foundation grant AST-2009210.
RES acknowledges support from the STFC (grant
number ST/P000525/1, ST/T000473/1). 
This work used the DiRAC@Durham
facility managed by the Institute for Computational Cosmology on
behalf of the STFC DiRAC HPC Facility (www.dirac.ac.uk). The equipment
was funded by BEIS capital funding via STFC capital grants
ST/K00042X/1, ST/P002293/1, ST/R002371/1 and ST/S002502/1, Durham
University and STFC operations grant ST/R000832/1. DiRAC is part of
the National eInfrastructure. 
This research made use of numpy, a
library used for scientific computing and technical computing and
matplotlib, a Python library for publication quality graphics
\citep{Harris2020, Hunter2007}.
\section*{Data Availability}
Instructions on how to access and download the data of the Quijote Simulation suite can be found in \url{https://github.com/franciscovillaescusa/Quijote-simulations}.  Additional data underlying this article will be shared on reasonable request to the corresponding author.
%
%
%
\bibliographystyle{mnras} \bibliography{refs} 


\appendix
\section{Details of the implementation}
\label{app:ExplicitExpressions}
\subsection{Choice of mass matrix}\label{sapp:MassMatrix}
The mass matrix can be written in
a block like structure consisting of the Hessian associated with
the latent field parameters, the Hessian with respect to the power
spectrum amplitude, and a mixed one. For this work we make the
following choices:
\begin{align} \mathbf{M} &= \bracmat
{\nabla^2_{\vec{\delta}}\left(\psi_{\mathrm{Poiss}}
+\psi_{\mathrm{Gauss}}\right)}
{\nabla^2_{\vec{\delta},\vec{\Pi}_{P^*}} \psi_{\mathrm{Gauss}}}
{\nabla^2_{\vec{\delta},\vec{\Pi}_{P^*}} \psi_{\mathrm{Gauss}}}
{\nabla^2_{\vec{\Pi}_{P^*}} \psi_{\mathrm{Gauss}}} \nonumber \\
&\equiv \bracmat {\nabla^2_{\vec{a}}\psi_{\mathrm{Gauss}} +
\vec{\varepsilon}} {-\frac{\partial \vec{P} /
\partial\vec{\Pi}_{P^*}}{\left({\vec{P}^*}\right)^{3/2}}} {-\frac{\partial \vec{P}
/ \partial\vec{\Pi}_{P^*}}{{\left(\vec{P}^*\right)}^{3/2}}} {\frac{1}{2}
\frac{\partial \vec{P} / \partial\vec{\Pi}_{P^*}}{\vec{P}^*} \cdot
\frac{\partial \vec{P} / \partial\vec{\Pi}_{P^*}}{\vec{P}^*}}
\nonumber \\ &= \bracmat {\frac{\mathds{1}}{\vec{P}^*}\left(\vec{1} +
\vec{P}^*\odot \vec{\vec{\varepsilon}}\right)}
{-\frac{1}{A^*\sqrt{\vec{P}^*}}}
{-\left(\frac{1}{A^*\sqrt{\vec{P}^*}}\right)^T} {\frac{n_L}{2{A^*}^2}}
\ ,
\end{align} where in the first step we drop the Poisson contribution,
introduce a positive definite, constant normalization vector
$\vec{\varepsilon}$ that is needed to keep the mass well defined, and
substitute the Fourier space representation of
\Eqn{eq:4_GaussianPrior} evaluated at $a^\dagger a = P^*$. In the second step we specialize to the case in which
$\vec{\Pi}_P^* \equiv A^*$ and introduce the symbol $\odot$ to denote the Hadamard product. As the latent part has a diagonal
structure we can easily compute the inverse with help of the block
matrix inversion formula:
\begin{align} \mathbf{M}^{-1} &= \bracmat
{\left(\sqrt{\vec{P}^*}\odot\vec{X}\right)\mathds{1} + \alpha
\vec{X}\otimes\vec{X}} {\alpha A^* \vec{X}} {\alpha A^* \vec{X}^T}
{\alpha {A^*}^2} \ \ \ \ , \nonumber \\ &{\mathrm{where}}
\hspace{.5cm} \vec{X} \equiv
\frac{\sqrt{\vec{P}^*}}{\vec{1}+\vec{P}^*\odot
\vec{\vec{\varepsilon}}} \hspace{.5cm} \alpha \equiv \frac{2}{n_L}
\frac{1}{1-2\left\langle\frac{1}{1+\vec{P}^*\odot\vec{\vec{\varepsilon}}}\right\rangle}
\ \ .
\end{align} To circumvent the large storage requirements of this
representation we make use the fact that we only need to perform the
operation $\mathbf{M}^{-1}\vec{p}$ within the integrator which can be
done in linear time and space complexity as
\begin{align} \mathbf{M}^{-1}\bracvec{\vec{p}_L}{\vec{p}_{\vec{\Pi}}} =
\bracvec {\sqrt{\vec{P}^*} \odot \vec{X}\odot\vec{p}_L + \alpha \Sigma
\vec{X} + \alpha A^* p_\Pi \vec{X}}{\alpha A^* \Sigma +{A^*}^2\alpha
p_\Pi} \ ,
\end{align} where we defined $\Sigma \equiv \vec{X}^T \cdot
\vec{p}_L$. We can now draw a random momentum vector as follows:
\begin{enumerate}
	\item Compute a lower triangular matrix $\mathbf{L}$
s.t. $\mathbf{M}=\mathbf{L}\mathbf{L}^T$. In our case the resulting Cholesky
matrix reads
	\begin{align} \mathbf{L} = \bracmat
{\frac{\mathds{1}}{\sqrt{\vec{P}^*}}\sqrt{\vec{1}+\vec{P}^*\odot\vec{\vec{\varepsilon}}}}
{0} {-\left(\frac{\vec{X}}{A^*\sqrt{\vec{P}^*}}\right)^T}
{\frac{1}{\sqrt{\alpha} A^*}} \ .
    \end{align}
	\item Draw a random unit Gaussian, $\vec{z} \sim
\mathcal{H}(0,\mathds{1}_{n_L}) \oplus \mathcal{G}(0, 1)$, where for
the latent field pixels we enforce hermitian symmetry.
	\item Transform $z$ to inherit the correct covariance
properties: $\vec{p} = \mathbf{L}\vec{z}$. This can again be done without
needing to store $\mathbf{L}$ as a whole.
\end{enumerate}
\begin{figure}
    \centering
    \centering{
    \includegraphics[width=.98\columnwidth]{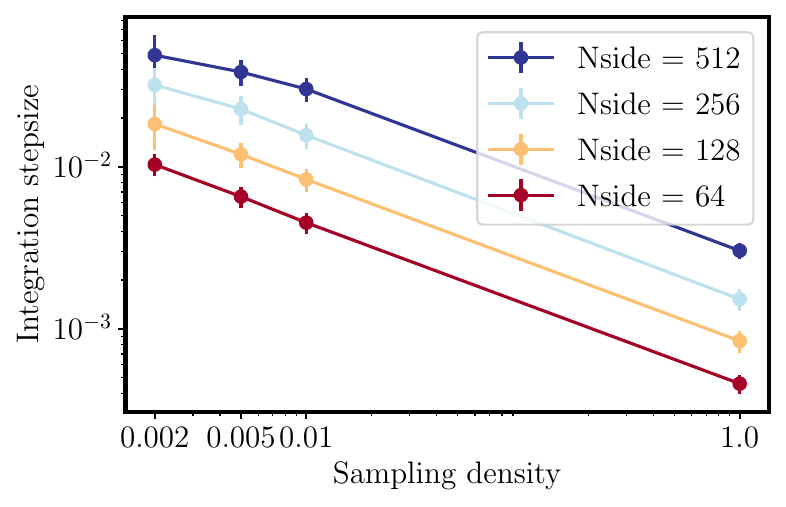}
    \includegraphics[width=.98\columnwidth]{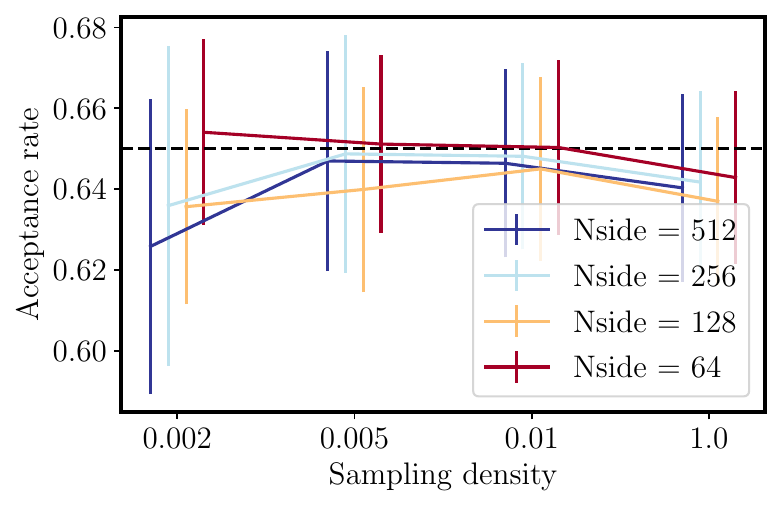}
    }
    
      \caption{{\textit{Upper panel}: The final step-size proposed by the dual averaging algorithm for the DoubleLog models. \textit{Lower panel}: Mean acceptance rate of the chains during the sampling stage. The black dashed line indicates the target acceptance rate of $\delta \equiv 0.65$ used within the dual averaging algorithm. For both plots we took into account all the chains in the DoubleLog models and averaged over all the seeds.}}
      \label{fig:DualAveraging}
\end{figure}
\subsection{Dual Averaging}\label{sapp:DualAveraging}
We implement a dual averaging algorithm using the same formalism as presented in \cite{HoffmanGelman2014}, see their Algorithm 5 for the specific implementation and initialization values. At its core, their dual averaging algorithm aims at dynamically adjusting the step-size $\epsilon$ of the chain during burn-in such that some target acceptance rate $\delta$ is reached, where $\delta$ should be chosen to maximize the trajectory length (i.e. lowering the correlation between subsequent samples) while keeping the rejection rate sufficiently low. Compared to traditional adaptation methods \citep{RobbinsMonro1951,AtchadeRosenthal2005} the dual averaging method does give a larger weight to more recent iterations and therefore allows for a quicker convergence to the 'optimal' step-size. As our target acceptance rate we chose $\delta = 0.65$ as proposed by \cite{HoffmanGelman2014}. In \Fig{fig:DualAveraging} we show how well this setup did work in the ensemble runs which were all starting the first burn-in stage with the same, very small step-size. We see that the resulting step-size is indeed compatible with the target acceptance rate for all the probed ensemble runs. We also see that there is a strong dependence of the inferred step-size on the galaxy sampling density and the pixel scale.
\begin{figure}
    \centering
    \centering{
    \includegraphics[width=1\columnwidth]{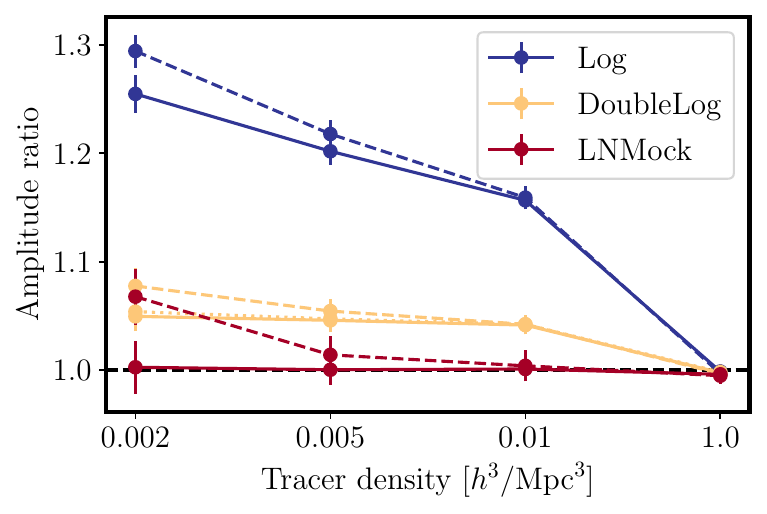}
    }
      \caption[Step-size dependent bias]{{Dependence of the power spectrum amplitude ratio on the target acceptance rate $\delta$ in the dual averaging algorithm for different transformations in the ensemble runs performed on a $256^2$ grid. The solid lines display the results for $\delta=0.95$ using the fourth order integrator while for the dashed  lines we choose $\delta=0.65$ and use the Leapfrog integration routine. The yellow dotted line corresponds to an ensemble run using the Leapfrog integrator with $\delta=0.95$.}}
      \label{fig:StepsizeBias}
\end{figure}
However, we also find that for this automatically inferred step-size there is a non-negligible amplitude bias present in the lognormal mocks for a very low galaxy sampling density. When using a much more restrictive value of $\delta = 0.95$ in the dual averaging algorithm we find that this bias disappears. In \Fig{fig:StepsizeBias} we compare the amplitude biases resulting from the different choices of $\delta$ for the Quijote snapshots and see that the most significant differences occur for small tracer densities. We note that the necessity of accepting nearly all proposed samples leads to the increased efficiency of the fourth order integrators compared to the Leapfrog routine as the former ones are not as strongly affected by the choice of $\delta$. Finally, we remark that the choice of $\delta$ does not significantly affect our estimate of the reconstruction confidence.
\subsection{Symplectic integrators}
In a more compact form the Hamilton  equations of motion \Eqn{eq:4_HamiltonianEOM} can be written in terms of the phase space variable $\vec{z}\equiv(\vec{q},\vec{p})$ and the Poisson bracket $\{\cdot,\cdot\}$ as 
\begin{align}
    \frac{\d}{\d t}\vec{z} \equiv -\{\mathcal{H},\vec{z}\}
    \ .
    \label{eq:HOM_compact}
\end{align}
We furthermore introduce the flow map $\phi_t$ which evolves some initial condition $\vec{z}_0$ along the Hamiltonian trajectory; the underlying geometry of Hamiltonian dynamics then demands that the flow map is a symplectomorphism such that we also need to embed the properties of this feature into our numerical approximation of the true flow. 

Treating the Poisson bracket as a differential operator, $\{X,Y\} \equiv D_Y X$, we can formally solve \Eqn{eq:HOM_compact} using an exponential:
\begin{align}
\vec{z}(t) = \exp(D_{\mathcal{H}}t) \ \vec{z}(0)
\equiv \exp[(D_\psi + D_T)t] \ \vec{z}(0) \ ,
\end{align}
where we split the operator in its Potential ($D_\psi$) and kinetic ($D_T$) parts. On their own, the individual summands update the phase space vector by discretely evolving (translating) the Hamiltonian equations of motion  \Eqn{eq:4_HamiltonianEOM} for a single  time step $\epsilon$:
\begin{align}
    \mathrm{e}^{D_T \epsilon} \equiv \mathcal{T}_{\vec{q}}(\epsilon)
    \ \ : \ \ 
    \left(\vec{q},\vec{p}\right) 
    \rightarrow
    \left(\vec{q}+\epsilon M^{-1}\vec{p},\vec{p}\right) 
    \nonumber \ , \\
    \mathrm{e}^{D_{\mathcal{\psi}} \epsilon} \equiv \mathcal{T}_{\vec{p}}(\epsilon)
    \ \ : \ \ 
    \left(\vec{q},\vec{p}\right) 
    \rightarrow
    \left(\vec{q},\vec{p}-\epsilon \nabla_{\vec{q}} \mathcal{\psi}\right) \ .
\end{align}
In order to construct an $n$th order explicit discretisation scheme we need to find an approximate solution up to an error $\mathcal{O}(t^n)$ in which the two differential operators are split:
\begin{align}
    \exp[(D_\psi + D_T)t+ \mathcal{O}(t^n)] = \prod_{i=1}^k \exp(c_iD_\psi t)\exp(d_i D_T t) + \mathcal{O}(t^n) \ .
\end{align}
Finding a set of algebraic requirement on the coefficients to guarantee an integrator of order $n$ can be achieved by making use of the Baker-Campbell-Haussdorff formula that can be repeatedly applied to yield:
\begin{align}
    \mathrm{e}^X \mathrm{e}^Y \mathrm{e}^X = \mathrm{e}^W
    \ : \ 
    W = 2X + Y + \frac{1}{6} [Y,[Y,X]] -\frac{1}{6}[X,[X,Y]] + \cdots
\end{align}
As the most simple example we can verify the discretisation scheme used in the Leapfrog integrator: \begin{align}
    \phi^{\mathrm{(lf)}}_\epsilon = \mathrm{e}^{c_1 D_\psi \epsilon} &\mathrm{e}^{d_1 D_T \epsilon} \mathrm{e}^{c_1D_\psi \epsilon}
    + \mathcal{O}(\epsilon^3)
    \equiv 
    \mathrm{e}^{(D_\psi+D_T)\epsilon+ \mathcal{O}(\epsilon^3)}
    \nonumber \\
    &\Rightarrow \ \ \ 
    \phi^{\mathrm{(lf)}}_\epsilon =
    \mathcal{T}_{\vec{p}}(\epsilon/2)\mathcal{T}_{\vec{q}}(\epsilon)\mathcal{T}_{\vec{p}}(\epsilon/2)
    \ .
\end{align}
\begin{figure}
    \centering
    \centering{
    \includegraphics[width=1\columnwidth]{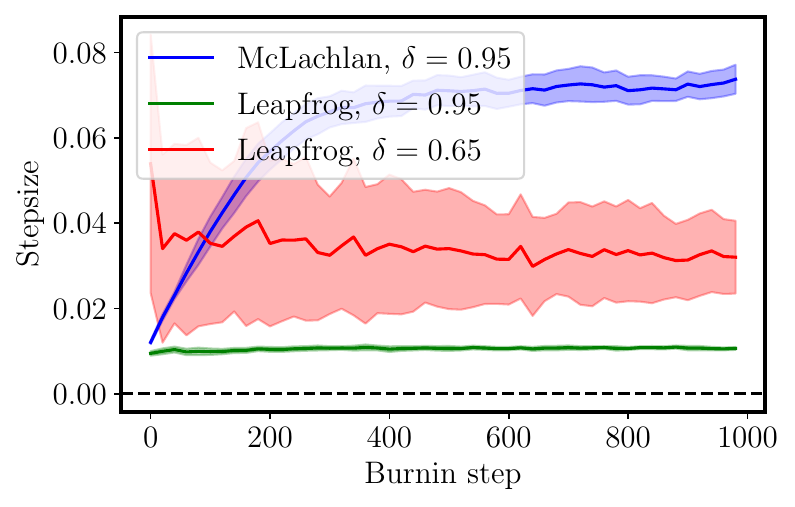}
    }
      \caption[Stepsize comparison]{Comparison of the integration step-size evolution during the burnin stage for the second and fourth order integrators across the corresponding ensembles using $256^2$ grids. As we only store thinned chains for those measurements the curves appear slightly discontinuous.}
      \label{fig:StepsizeComparison}
\end{figure}
\begin{figure*}
    \centering
    \centering{
    \includegraphics[width=.98\columnwidth]{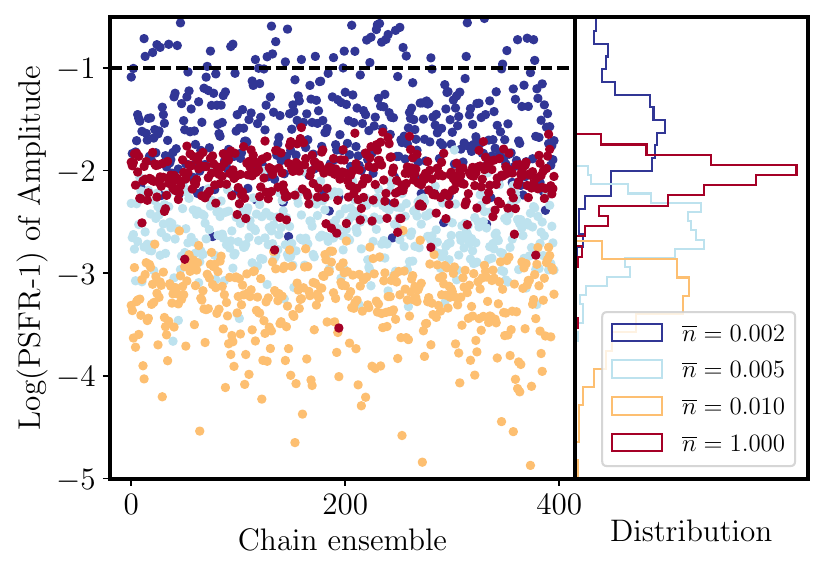}
    \includegraphics[width=.98\columnwidth]{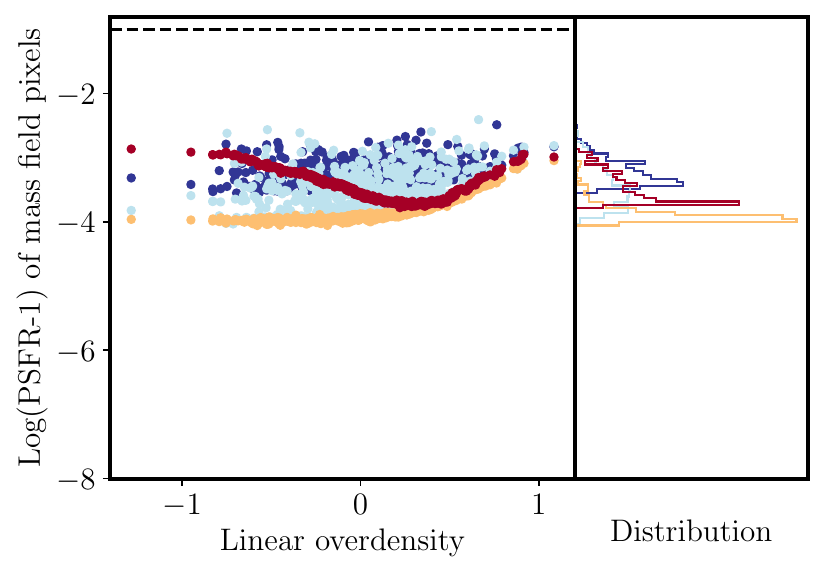}
    \\
    \includegraphics[width=.98\columnwidth]{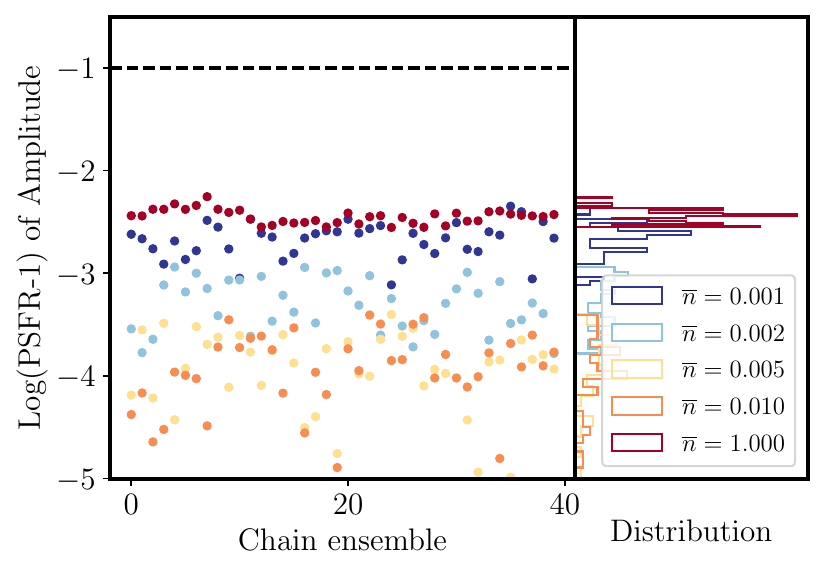}
    \includegraphics[width=.98\columnwidth]{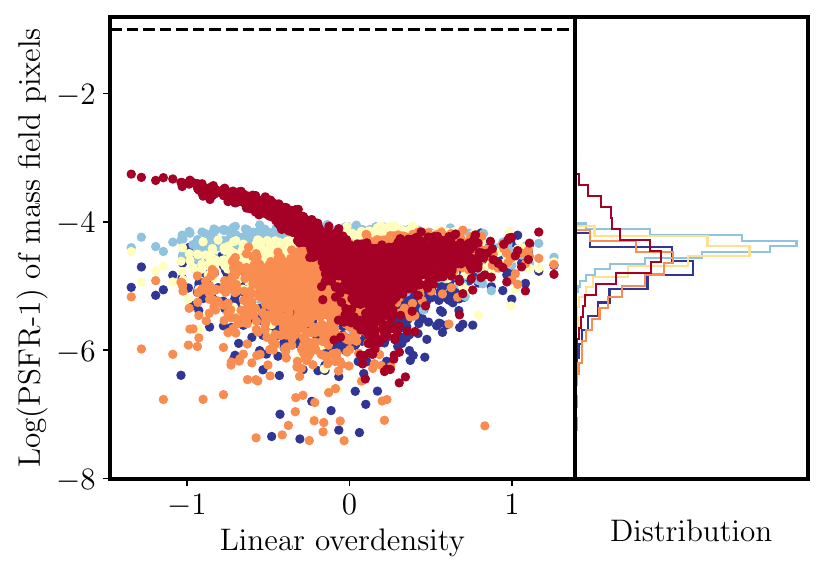}
    }
      \caption[Convergence of the chains in the Quijote baseline ensemble run]{Convergence diagnostic of the chains in the Quijote baseline ensemble run (top) and the corresponding run using the fourth order integrator (bottom). The left hand side shows the results for the power spectrum amplitude for each of the individual runs. The color coding corresponds to the different galaxy sampling densities. On the right hand side we plot the ensemble mean of the PSFR for $500$ (top) and $2000$ (bottom) randomly chosen pixels. In each panel the dashed black line indicates the boundary of our chosen measure of convergence; point below this line signify a set of chains that have converged.}
      \label{fig:PSFRs}
\end{figure*}
For obtaining higher order integrators one needs to chain together more individual updates. In order to find locally optimal solutions for the coefficients one usually opts for sufficiently many updates, such that in addition to the algebraic constraints the remaining coefficients can be chosen by optimizing certain quantities, such as leading order terms in the error expansion.
In particular, the fourth order integrator of \cite{McLachlan1995} is obtained by symmetrically concatenating five Leapfrog routines using different weights:
\begin{align}
    \phi^{\mathrm{(mcl)}}_{\epsilon}
    =
    \phi^{\mathrm{(lf)}}_{w_1\epsilon} \circ
    \phi^{\mathrm{(lf)}}_{w_2\epsilon} \circ
    \phi^{\mathrm{(lf)}}_{w_3\epsilon} \circ
    \phi^{\mathrm{(lf)}}_{w_2\epsilon} \circ
    \phi^{\mathrm{(lf)}}_{w_1\epsilon} \ ,
\end{align}
where the optimal weights are given as $w_1=0.28$, $w_2=0.62546642846767004501$ and $w_3=1-2(w_1+w_2)$. A naive implementation of this method will require $15$ single updates per integration step. However, we can reduce this number by making use of the fact that the position vector is not affected in the momentum update; this allows us to concatentate the momentum updates in between adjacent leapfrog steps.
Note that when evolving an initial state for $m > 1$ time steps $\epsilon$ it is possible to further reduce the number of operations by additionally chaining together the final momentum updates in between steps. If an integrator $x$ requires $n_x^{(1)}$ stages for a single time step the full evolution can be obtained using 
\begin{align*}
    n_x^{(m)} = n_x^{(1)} + (m-1)\left( n_x^{(1)} -1 \right)
    \overset{m \gg 1}{\rightarrow} m\left( n_x^{(1)} -1 \right)
\end{align*}
single updates. For the Leapfrog integrator and the fourth integrator of \citep{McLachlan1995} we have $n_{\mathrm{lf}}^{(1)}=3$ and $n_{\mathrm{mcl}}^{(1)}=11$ such that the fourth order integrator asymptotes to requiring five times more time than the leapfrog routine. In \Fig{fig:StepsizeComparison} we show the difference of the inferred step-size for the Leapfrog and fourth order integrator and we infer that for such model specifications a speedup of around $50\%$ can be expected during the sampling phase.

We postpone a more thorough investigation of the applicability and performance of various other higher order integration schemes to future work. For additional details on numerical Hamiltonian dynamics we refer the reader to \cite{Leimkuhler2004}.
\section{Convergence tests}
In \Fig{fig:PSFRs} we show the potential scale reduction factor (PSFR) of the Gelman-Rubin diagnostics \citep{GelmanRubin1992} for the DoubleLog$\_$BaseRes and the DoubleLog$\_$BaseResrun$\_$mcl runs. We see that the amplitude chains did converge for practically all the runs, except for some reconstructions performed with very low galaxy sampling densities in the leapfrog case. However, as those PSFR values are only slightly above the $1.1$ level and we just care about the ensemble mean of the distributions, this should not pose a problem. For the latent pixel parameters we see that the mean PSFR for all the runs is very well consistent with unity which signifies convergence of the individual chains. Performing Gelman-Rubin diagnostics on the other ensemble runs we find similar results.

\section{Correlation matrices of the projected fields}
\label{app:CorrelationMatrix}
\begin{figure*}
    \centering
    \centering{
    \includegraphics[width=.99\columnwidth]{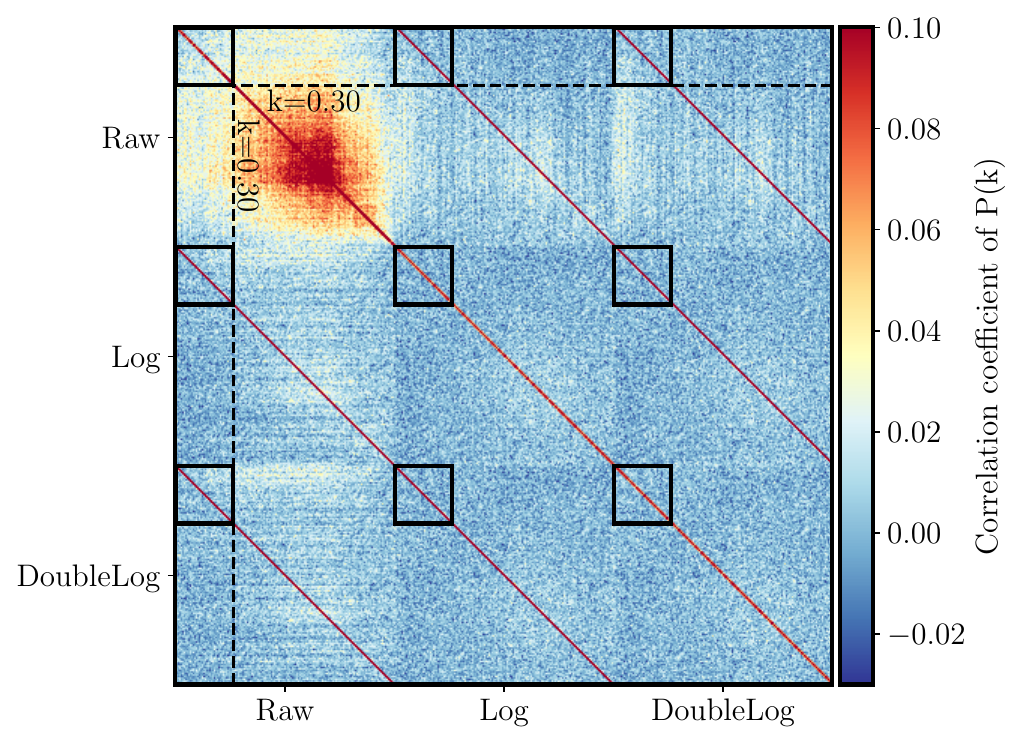}
    \includegraphics[width=.99\columnwidth]{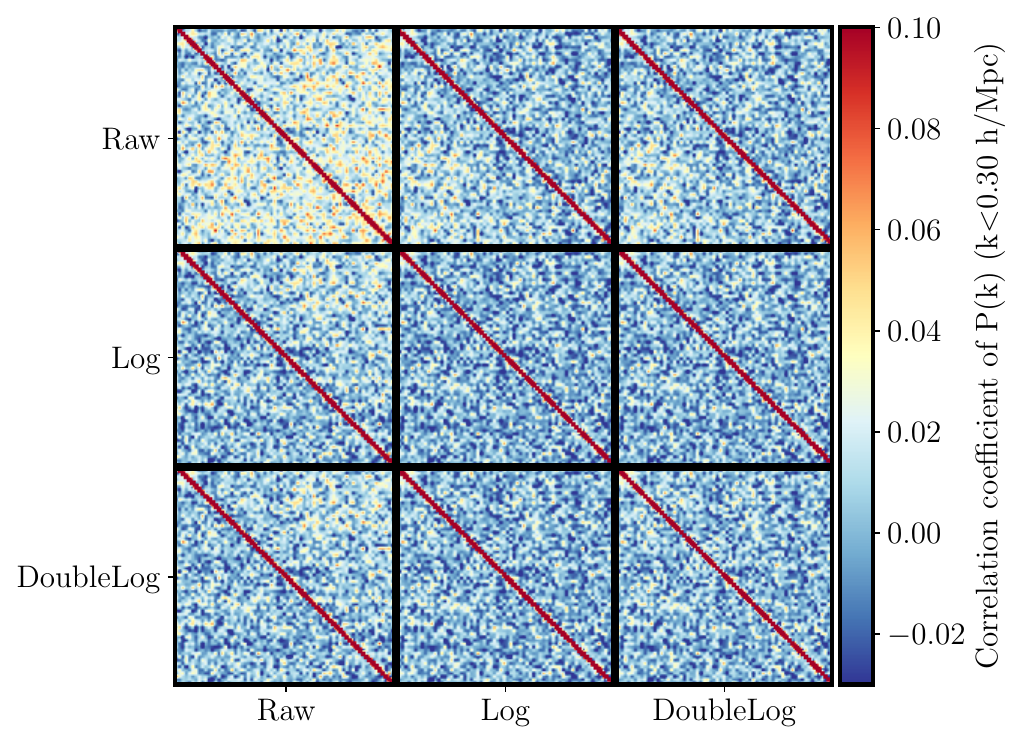}
    }
      \caption{\textit{Left hand side:} Correlation matrices of the power spectra for the nonlinear field and its two transformations used in this work (transverse resolution $\approx 4 \ h^{-1}$Mpc ). \textit{Right hand side:} Correlation matrix of the blocks with $k<0.3 \ h$ Mpc$^{-1}$. For those scales there is only a slight amount of mode coupling present in the spectra of the raw field s.t. the information content will be only depleted by a bit with respect to the Gaussian limit.}
      \label{fig:Corrcoefs}
\end{figure*}

In \Fig{fig:Corrcoefs} we show the correlation structure of the power spectra of the three projected fields discussed in this work, namely the untransformed field, the log-transformed field and the DoubleLog-transformed field for their base resolution of $\approx 4~h^{-1}$Mpc. We see that on small scales ($k \gtrsim 0.3 \ h$Mpc$^{-1}$) there is significant mode coupling in the spectra of the untransformed field while the correlation matrix of the spectra of the transformed fields exhibit a nearly diagonal structure. Thus, we expect the information content of the raw field to deplete relative to a Gaussian field at those scales and to reach a plateau in the high-$k$ regime. In contrast, for both of the transformed fields there is no significant mode coupling at any scale probed in this work such that these fields contain a similar amount information as a Gaussian field.

\bsp	
\label{lastpage}
\end{document}